\documentclass{pnastwo}

\usepackage{amssymb,amsfonts,amsmath}
\usepackage{natbib}
\usepackage[pdftex]{graphicx}
\usepackage{algorithm,algorithmic}

\newcommand{\bu}{\mathbf{u}}

\newcommand{\by}{\mathbf{y}}
\newcommand{\bz}{\mathbf{z}}

\renewcommand{\vss}{\vspace{0.25cm}}
\newcommand{\btheta}{\boldsymbol{\theta}}
\newcommand{\feta}{\boldsymbol{\eta}}

\url{www.pnas.org/cgi/doi/10.1073/pnas.xxx}
\copyrightyear{2011}
\issuedate{Issue Date}
\volume{Volume}
\issuenumber{Issue Number}

\begin{document}
\title{Lack of confidence in ABC model choice}

\author{
Christian P.~Robert,\affil{1}{Universit\'e Paris Dauphine}\affil{2}{Institut Universitaire de France}\affil{3}{CREST, Paris}
Jean-Marie Cornuet,\affil{4}{CBGP, INRA, Montpellier, France}
Jean-Michel Marin,\affil{5}{I3M, UMR CNRS 5149, Universit\'e Montpellier 2, France}
\and
Natesh S.~Pillai\affil{6}{Department of Statistics, Harvard University, Cambridge, MA}
}

\maketitle
\begin{article} 

\begin{abstract}
Approximate Bayesian computation (ABC) have become an essential tool for the analysis of complex stochastic
models.  Grelaud et al. (2009, Bayesian Ana 3:427--442) advocated the use of ABC for model choice in the
specific case of Gibbs random fields, relying on a inter-model sufficiency property to show that the
approximation was legitimate. We implemented ABC model choice in a wide range of phylogenetic models in the
DIY-ABC software (Cornuet et al. (2008) Bioinfo 24:2713--2719). We now present arguments as to why the
theoretical arguments for ABC model choice are missing, since the algorithm involves an unknown loss of
information induced by the use of insufficient summary statistics. The approximation error of the posterior
probabilities of the models under comparison may thus be unrelated with the computational effort spent in
running an ABC algorithm.  We then conclude that additional empirical verifications of the performances of the
ABC procedure as those available in DIYABC are necessary to conduct model choice.
\footnote{CPR, JMM and NSP designed and performed research, JMC and JMM analysed data, and CPR, JMC and JMM
wrote the paper.} \end{abstract} 
\keywords{likelihood-free methods | Bayes factor | Bayesian model choice | sufficiency}

\abbreviations{ABC, approximate Bayesian computation; ABC-MC, ABC model choice; DIY-ABC, Do-it-yourself ABC;
IS, importance sampling; SMC, sequential Monte Carlo}

\dropcap{I}nference on population genetic models such as coalescent trees is one representative example of
cases when statistical analyses like Bayesian inference cannot easily operate because the likelihood function
associated with the data cannot be computed in a manageable time
\citep{tavare:balding:griffith:donnelly:1997,beaumont:zhang:balding:2002,cornuet:santos:beaumont:etal:2008}.
The fundamental reason for this impossibility is that the model associated with coalescent data has to
integrate over trees of high complexity.

In such settings, traditional approximation tools like Monte Carlo simulation \citep{robert:casella:2004} from
the posterior distribution are unavailable for practical purposes. Indeed, due to the complexity of the latent
structures defining the likelihood (like the coalescent tree), their simulation is too unstable to bring a
reliable approximation in a manageable time. Such complex models call for a practical if cruder approximation
method, the ABC methodology
\citep{tavare:balding:griffith:donnelly:1997,pritchard:seielstad:perez:feldman:1999}.  This rejection technique
bypasses the computation of the likelihood via simulations from the corresponding distribution (see
\citealp{beaumont:2010} and \citealp{lopes:beaumont:2010} for recent surveys, and
\citealp{csillery:blum:gaggiotti:francois:2010} for the wide and successful array of applications based on
implementations of ABC in genomics and ecology). 

We argue here that ABC is a generally valid approximation method for doing Bayesian inference in complex
models. However, without further justification, ABC methods cannot be trusted to discriminate between two
competing models when based on insufficient summary statistics. We exhibit simple examples in which the
information loss due to insufficiency leads to inconsistency, i.e.~when the ABC model selection fails to
recover the true model, even with infinite amounts of observation and computation. On the one hand, ABC using
the entire data leads to a consistent model choice decision but it is clearly infeasible in most settings. On
the other hand, too much information loss due to insuffiency leads to a statistically invalid decision
procedure. The challenge is in achieving a balance between information loss and consistency.  Theoretical
results that mathematically validate model choice for insufficient statistics are currently lacking on a
general basis.

Our conclusion at this stage is to opt for a cautionary approach in ABC model choice, handling it as an
exploratory tool rather than trusting the Bayes factor approximation. The corresponding degree of approximation
cannot be evaluated, except via Monte Carlo evaluations of the model selection performances of ABC.  More
empirical measures such as those proposed in the DIY-ABC software \citep{cornuet:santos:beaumont:etal:2008} and
in \cite{ratmann:andrieu:wiujf:richardson:2009} 
thus seem to be the only available solution at the current time for conducting model comparison. 

We stress that, while \cite{templeton:2008,templeton:2010} repeatedly expressed reservations about the formal
validity of the ABC approach in statistical testing, those criticisms were 
rebutted in \cite{clade:2010,csillery:blum:gaggiotti:francois:2010b,berger:fienberg:raftery:robert:2010} and
are not relevant for the current paper. 

\section{Statistical Methods}

\subsection{The ABC algorithm}\label{Genesis}
The setting in which ABC operates is the approximation of a simulation from the posterior distribution
$\pi(\btheta|\by) \propto \pi(\theta) f(\by|\btheta)$ when distributions associated with both the prior $\pi$
and the likelihood $f$ can be simulated (the later being unavailable in closed form).  The first ABC algorithm was introduced by
\cite{pritchard:seielstad:perez:feldman:1999} 
as follows: given a sample $\by$ from a sample space $\mathcal{D}$, a sample $(\theta_1,\ldots,\theta_M)$ is produced by\\

\noindent{\bf Algorithm 1: ABC sampler}
\begin{algorithmic}\label{algo:ABC0}
\FOR {$i=1$ to $N$}
\REPEAT
\STATE Generate $\btheta'$ from the prior distribution $\pi(\cdot)$
\STATE Generate $\bz$ from the likelihood $f(\cdot|\btheta')$
\UNTIL {$\rho\{\eta(\bz),\eta(\by)\}\leq \epsilon$}
   \STATE set $\btheta_i=\btheta'$,
\ENDFOR
\end{algorithmic}

\smallskip
The parameters of the ABC algorithm are the so-called summary statistic $\eta(\cdot)$, the distance
$\rho\{\cdot,\cdot\}$, and the tolerance level $\epsilon>0$. The approximation of the posterior distribution
$\pi(\btheta|\by)$ provided by the ABC sampler is to instead sample from the marginal in $\btheta$ of the joint
distribution
$$
\pi_\epsilon(\btheta,\bz|\by)=
\frac{\pi(\btheta)f(\bz|\btheta)\mathbb{I}_{A_{\epsilon,\by}}(\bz)}
{\int_{A_{\epsilon,\by}\times\Theta}\pi(\btheta)f(\bz|\btheta)\text{d}\bz\text{d}\btheta}\,,
$$
where $\mathbb{I}_B(\cdot)$ denotes the indicator function of $B$ and 
$$
A_{\epsilon,\by}=\{\bz\in\mathcal{D}|\rho\{\eta(\bz),\eta(\by)\}\leq \epsilon\} \,.
$$
The basic justification of the ABC approximation is that, when using a sufficient statistic $\eta$ and 
a small (enough) tolerance $\epsilon$, we have 
$$
\pi_\epsilon(\btheta|\by)=\int \pi_\epsilon(\btheta,\bz|\by)\text{d}\bz\approx \pi(\btheta|\by)\,.
$$

In practice, the statistic $\eta$ is necessarily insufficient (since only exponential families enjoy sufficient
statistics with fixed dimension, see \citealp{lehmann:casella:1998}) and the approximation then converges to
the less informative $\pi(\btheta|\eta(\by))$ when $\epsilon$ goes to zero. This loss of information is a
necessary price to pay for the access to computable quantities and $\pi(\btheta|\eta(\by))$ provides a
convergent inference on $\btheta$ when $\btheta$ is identifiable in the distribution of $\eta(\by)$
\citep{fearnhead:prangle:2010}. While acknowledging the gain brought by ABC in handling Bayesian inference in
complex models, and the existence of involved summary selection mechanisms
\citep{joyce:marjoram:2008,nunes:balding:2010}, we demonstrate here that the loss due to the ABC approximation
may be arbitrary in the specific setting of Bayesian model choice via posterior model probabilities.

\subsection{ABC model choice}
The standard Bayesian tool for model comparison is the marginal likelihood \citep{jeffreys:1939}
$$
w(\by) = \int_\Theta \pi(\btheta) f(\by|\btheta)\,\text{d} \btheta\,,
$$
which leads to the Bayes factor for comparing the evidences of models with likelihoods 
$f_1(\by|\btheta_1)$ and $f_2(\by|\btheta_2)$,
$$
B_{12}(\by) = \dfrac{w_1(\by)}{w_2(\by)} =\dfrac{\int_{\Theta_1} \pi_1(\btheta_1) f_1(\by|\btheta_1)\,\text{d}
\btheta_1}{\int_{\Theta_2} \pi_2(\btheta_2) f_2(\by|\btheta_2)\,\text{d} \btheta_2}\,.
$$
As detailed in \cite{clade:2010}, it provides a valid criterion
for model comparison that is naturally penalised for model complexity.

Bayesian model choice proceeds by creating a probability structure across $M$ models (or likelihoods). 
It introduces the model index $\mathcal{M}$ as an extra unknown parameter, associated with its prior distribution, $\pi(\mathcal{M}=m)$
($m=1,\ldots,M$), while the prior distribution on the parameter is conditional on the value $m$ of the $\mathcal{M}$
index, denoted by $\pi_m(\btheta_m)$ and defined on the parameter space $\Theta_m$.  The choice between those
models is then driven by the posterior distribution of $\mathcal{M}$, 
$$
\mathbb{P}( \mathcal{M}=m | \by ) = {\pi(\mathcal{M}=m) w_m(\by) }\big/ {\sum_k \pi(\mathcal{M}=k) w_k(\by) }
$$
where $w_k(\by)$ denotes the marginal likelihood for model $k$.

While this posterior distribution is straightforward to interpret, it offers a challenging computational
conundrum in Bayesian analysis.  When the likelihood is not available, ABC represents the almost unique
solution.  \cite{pritchard:seielstad:perez:feldman:1999} describe the use of model choice based on ABC for
distinguishing between different mutation models.  The justification behind the method is that the average ABC
acceptance rate associated with a given model is proportional to the posterior probability corresponding to this
approximative model, when identical summary statistics, distance, and tolerance level are used over all models.
In practice, an estimate of the ratio of marginal likelihoods is given by the ratio of observed acceptance
rates.  Using Bayes formula, estimates of the posterior probabilities are straightforward to derive.  This
approach has been widely implemented in the literature (see, e.g., \citealp{estoup:etal:2004},
\citealp{miller:etal:2005}, \citealp{pascual:etal:2007}, and \citealp{sainudiin:etal:2011}). 

A representative illustration of the use of an ABC model choice approach is given by
\cite{miller:etal:2005} which analyses the European invasion of the western corn rootworm, North America's
most destructive corn pest. Because this pest was initially introduced in Central Europe, it was believed that
subsequent outbreaks in Western Europe originated from this area.  Based on an ABC model choice analysis of
the genetic variability of the rootworm, the authors conclude that this belief is false: There have been at
least three independent introductions from North America during the past two decades.

The above estimate is improved by regression regularisation \citep{fagundes:etal:2007},
where model indices are processed as categorical variables in a 
polychotomous regression. When comparing two models, this involves a standard logistic regression.
Rejection-based approaches were lately introduced by \cite{cornuet:santos:beaumont:etal:2008},
\cite{grelaud:marin:robert:rodolphe:tally:2009} and \cite{toni:etal:2009}, in a Monte Carlo 
simulation of model indices as well as model parameters. Those recent extensions are already widely used
in population genetics, as exemplified by
\cite{belle:etal:2008,cornuet:ravigne:estoup:2010,excoffier:leuenberger:wegman:2009,ghirotto:etal:2010,guillemaud:etal:2009,leuenberger:wegmann:2010, patin:etal:2009,ramakrishnan:hadly:2009,verdu:etal:2009,wegmann:excoffier:2010}. 
Another illustration of the popularity of this approach is given by the availability of four
softwares implementing ABC model choice methodologies: 
\begin{itemize}
\item ABC-SysBio,
which relies on a SMC-based ABC for inference in system biology, including model-choice \citep{toni:etal:2009}.
\item ABCToolbox which proposes SMC and MCMC implementations, as well as Bayes factor approximation
\citep{wegman:etal:2011}.
\item DIYABC,
which relies on a regularised ABC-MC algorithm on population history using molecular markers
\citep{cornuet:santos:beaumont:etal:2008}.
\item PopABC,
which relies on a regular ABC-MC algorithm for genealogical simulation \citep{lopes:balding:beaumont:2009}.
\end{itemize}
As exposed in e.g. \cite{grelaud:marin:robert:rodolphe:tally:2009}, \cite{toni:stumpf:2010}, or
\cite{didelot:everitt:johansen:lawson:2011},
once $\mathcal{M}$ is incorporated within the parameters, the ABC approximation to its
posterior follows from the same principles as in regular ABC.  The corresponding implementation is as follows,
using for the summary statistic a statistic $\feta(\bz)=\{\eta_1(\bz),\ldots,\eta_M(\bz)\}$ that is the
concatenation of the summary statistics used for all models (with an obvious elimination of duplicates).\\

\noindent{\bf Algorithm 2: ABC-MC}
\begin{algorithmic}\label{algo:ABCMoC} 
\FOR {$i=1$ to $N$} 
\REPEAT 
\STATE Generate $m$ from the prior $\pi(\mathcal{M}=m)$ 
\STATE Generate $\btheta_{m}$ from the prior $\pi_{m}(\btheta_m)$ 
\STATE Generate $\bz$ from the model $f_{m}(\bz|\btheta_{m})$ 
\UNTIL {$\rho\{\feta(\bz),\feta(\by)\}\leq\epsilon$} 
\STATE Set $m^{(i)}=m$ and $\btheta^{(i)}=\btheta_m$ 
\ENDFOR 
\end{algorithmic} 

\smallskip
The ABC estimate of the posterior probability $\pi(\mathcal{M}=m|\by)$ is then the frequency of acceptances
from model $m$ in the above simulation
$ \hat{\pi}( \mathcal{M}=m | \by ) = N^{-1}\,\sum_{i=1}^N \mathbb{I}_{m^{(i)}=m}\,.  $
This also corresponds to the frequency of simulated pseudo-datasets from model $m$ that are closer to the data
$\by$ than the tolerance $\epsilon$. In order to improve the estimation by smoothing,
\cite{cornuet:santos:beaumont:etal:2008} follow the rationale that motivated the use of a local linear
regression in \cite{beaumont:zhang:balding:2002} and rely on a weighted polychotomous regression to
estimate $\pi(\mathcal{M}=m|\by)$ based on the ABC output. This modelling is implemented in the DIYABC software.

\subsection{The difficulty with ABC-MC}\label{TheRamones}

There is a fundamental discrepancy between the genuine Bayes factors/posterior
probabilities) and the approximations resulting from ABC-MC.

The ABC approximation to a Bayes factor, $B_{12}$ say, resulting from Algorithm 2 is
$$
\widehat{B_{12}}(\by) = {\pi(\mathcal{M}=2)} {\sum_{i=1}^N \mathbb{I}_{m^{(i)}=1}}\bigg/
{\pi(\mathcal{M}=1)}{\sum_{i=1}^N \mathbb{I}_{m^{(i)}=2}} \,.
$$
An alternative representation is given by
$$
\widehat{B_{12}}(\by) = \dfrac{\pi(\mathcal{M}=2)}{\pi(\mathcal{M}=1)}\,\dfrac{\sum_{t=1}^T 
\mathbb{I}_{m^{t}=1}\,\mathbb{I}_{\rho\{\feta(\bz^t),\feta(\by)\}\le \epsilon}}{\sum_{t=1}^T
\mathbb{I}_{m^{t}=2}\,\mathbb{I}_{\rho\{\feta(\bz^t),\feta(\by)\}\le \epsilon}}\,,
$$
where the pairs $(m^t,z^t)$ are simulated from the joint prior and $T$ is the number of simulations
necessary for $N$ acceptances in Algorithm 2. In order to study the limit of this
approximation, we first let $T$ go to infinity. (For simplification purposes and without loss of generality, we
choose a uniform prior on the model index.) The limit of $\widehat{B_{12}}(\by) $ is then
\begin{eqnarray*}
B_{12}^\epsilon(\by) &=& \dfrac{\mathbb{P}[\mathcal{M}=1,\rho\{\feta(\bz),\feta(\by)\} \le \epsilon]}
                             {\mathbb{P}[\mathcal{M}=2,\rho\{\feta(\bz),\feta(\by)\} \le \epsilon]}\\
                     &=& \dfrac{\iint \mathbb{I}_{\rho\{\feta(\bz),\feta(\by)\} \le \epsilon} 
\pi_1(\btheta_1)f_1(\bz|\btheta_1)\,\text{d}\bz\,\text{d}\btheta_1}
                             {\iint \mathbb{I}_{\rho\{\feta(\bz),\feta(\by)\} \le \epsilon} 
\pi_2(\btheta_2)f_2(\bz|\btheta_2)\,\text{d}\bz\,\text{d}\btheta_2}\\
                     &=& \dfrac{\iint \mathbb{I}_{\rho\{\feta,\feta(\by)\} \le \epsilon} 
\pi_1(\btheta_1)f_1^{\feta}(\feta|\btheta_1)\,\text{d}\feta\,\text{d}\btheta_1}
                             {\iint \mathbb{I}_{\rho\{\feta,\feta(\by)\} \le \epsilon} 
\pi_2(\btheta_2)f_2^{\feta}(\feta|\btheta_2)\,\text{d}\feta\,\text{d}\btheta_2}\,,
\end{eqnarray*}
where $f_1^{\feta}(\feta|\btheta_1)$ and $f_2^{\feta}(\feta|\btheta_2)$ denote the densities of $\feta(\bz)$ when
$\bz\sim f_1(\bz|\btheta_1)$ and $\bz\sim f_2(\bz|\btheta_2)$, respectively. By L'Hospital formula, if $\epsilon$
goes to zero, the above converges to
$$
B^{\feta}_{12}(\by)={\int \pi_1(\btheta_1)f_1^{\feta}(\feta(\by)|\btheta_1)\,\text{d}\btheta_1}\bigg/
                          {\int \pi_2(\btheta_2)f_2^{\feta}(\feta(\by)|\btheta_2)\,\text{d}\btheta_2}\,,
$$
namely the Bayes factor for testing model $1$ versus model $2$ based on the sole observation of
$\feta(\by)$. This result reflects the current perspective on ABC: the inference derived from
the ideal ABC output when $\epsilon=0$ only uses the information contained in $\feta(\by)$. Thus, in the
limiting case, i.e.~when the algorithm uses an infinite computational power, the ABC odds ratio does not
account for features of the data other than the value of $\feta(\by)$, which is why the limiting Bayes
factor only depends on the distribution of $\feta$ under both models. 

When running ABC for point estimation, the use of an insufficient statistic does not usually jeopardise
convergence of the method. As shown, e.g., in \cite[][Theorem 2]{fearnhead:prangle:2010}, the noisy version of
ABC as an inference method is convergent under usual regularity conditions for model-based Bayesian inference
\citep{bernardo:smith:1994}, including identifiability of the parameter for the insufficient statistic $\feta$.
In contrast, the loss of information induced by $\feta$ may seriously impact model-choice Bayesian inference.
Indeed, the information contained in $\feta(\by)$ is lesser than the information contained in
$\by$ and this even in most cases when $\feta(\by)$ is a sufficient statistic for {\em both models}. In other words,
{\em $\feta(\by)$ being sufficient for both $f_1(\by|\btheta_1)$ and $f_2(\by|\btheta_2)$ does not usually
imply that $\feta(\by)$ is sufficient for $\{m,f_m(\by|\btheta_m)\}$.} To see why this is the case, consider
the most favourable case, namely when $\feta(\by)$ is a sufficient statistic for both models. We then have by
the factorisation theorem \citep{lehmann:casella:1998} that $f_i(\by|\btheta_i) = g_i(\by)
f_i^{\feta}(\feta(\by)|\btheta_i)$ $(i=1,2)$, i.e.
\begin{eqnarray}\label{eq:DireStraits}
B_{12}(\by) &=& \dfrac{w_1(\by)}{w_2(\by)}
= \dfrac{\int_{\Theta_1} \pi(\btheta_1) g_1(\by) f_1^{\feta}(\feta(\by)|\btheta_1)\,\text{d}
\btheta_1}{\int_{\Theta_2} \pi(\btheta_2) g_2(\by) f_2^{\feta}(\feta(\by)|\btheta_2)\,\text{d} \btheta_2}\nonumber\\
&=& \dfrac{g_1(\by)\,\int \pi_1(\btheta_1)f_1^{\feta}(\feta(\by)|\btheta_1)\,\text{d}\btheta_1}
{g_2(\by)\,\int \pi_2(\btheta_2)f_2^{\feta}(\feta(\by)|\btheta_2)\,\text{d}\btheta_2}\nonumber\\ 
&=& \dfrac{g_1(\by)}{g_2(\by)}\,B^{\feta}_{12}(\by)\,.
\end{eqnarray}
Thus, unless $g_1(\by)=g_2(\by)$, as in the special case of Gibbs random fields detailed below,
the two Bayes factors differ by the ratio $g_1(\by)/g_2(\by)$, which is only
equal to one in a very small number of known cases. This decomposition is a straightforward
proof that a model-wise sufficient statistic is usually not sufficient across models, hence for model comparison.
An immediate corollary is that the ABC-MC approximation does not always converge to the exact Bayes factor.

The discrepancy between limiting ABC and genuine Bayesian inferences does not come as a surprise, because ABC
is indeed an approximation method. Users of ABC algorithms are therefore prepared for some degree of
imprecision in their final answer, a point stressed by \cite{wilkinson:2008} and \cite{fearnhead:prangle:2010}
when they qualify ABC as exact inference on a wrong model.  However, the magnitude of the difference between
$B_{12}(\by)$ and $B^{\feta}_{12} (\by)$ expressed by \eqref{eq:DireStraits} is such that there is no direct
connection between both answers. In a general setting, if $\feta$ has the same dimension as one component of
the $n$ components of $\by$, the ratio $g_1(\by)/g_2(\by)$ is equivalent to a density ratio for a sample of
size $\text{O}(n)$, hence it can be arbitrarily small or arbitrarily large when $n$ grows. Contrastingly, the
Bayes factor $B^{\feta}_{12}(\by)$ is based on an equivalent to a single observation, hence does not
necessarily converge with $n$ to the correct limit, as shown by the Poisson and normal examples below and in
SI. The conclusion derived from the ABC-based Bayes factor may therefore completely differ from the conclusion
derived from the exact Bayes factor and there is no possibility of a generic agreement between both, or even of
a manageable correction factor. This discrepancy means that a theoretical validation of the ABC-based model
choice procedure is currently missing and that, due to this absence, potentialy costly simulation-based
assessments are required when calling for this procedure.

Therefore, users must be warned that ABC approximations to Bayes factors do not perform as standard numerical or
Monte Carlo approximations, with the exception of Gibbs random fields detailed in the next section.
In all cases when $g_1(\by)/g_2(\by)$ differs from one, no inference on the true Bayes factor can be derived
from the ABC-MC approximation without further information on the ratio $g_1(\by)/g_2(\by)$, most often
unavailable in settings where ABC is necessary.

\cite{didelot:everitt:johansen:lawson:2011} also derived this relation between both Bayes factors in their
formula {\bf [18]}. While they still advocate the use of ABC model choice in the absence of sufficient
statistic, we stress that no theoretical guarantee can be given on the validity of the ABC approximation to the
Bayes factor and hence of its use as a model choice procedure.

Note that \cite{sousa:etal:2009} resort to full allelic distributions in an ABC framework, instead of chosing
summary statistics. They show how to apply ABC using allele frequencies to draw inferences in cases where
selecting suitable summary statistics is difficult (and where the complexity of the model or the size of
dataset prohibits to use full-likelihood methods). In such settings, ABC-MC does not suffer from the divergence
exhibited here because the measure of distance does not involve a reduction of the sample. The same comment
applies to the ABC-SysBio software of \cite{toni:etal:2009}, which relies on the whole dataset. The theoretical
validation of ABC inference in hidden Markov models by \cite{dean:singh:jasra:peters:2011} should also extend
to the model choice setting because the approach does not rely on summary statistics but instead on the whole
sequence of observations.

\section{Results} 
\subsection{The specific case of Gibbs random fields}\label{TheClash} 
In an apparent contradiction with the above, \cite{grelaud:marin:robert:rodolphe:tally:2009} showed that
the computation of the posterior probabilities of Gibbs random fields under competition can be
done via ABC techniques, which provide a converging approximation to the true Bayes factor. The reason
for this result is that, for these models in the above ratio \eqref{eq:DireStraits},
$g_1(\by)=g_2(\by)$.  The validation of an ABC comparison of Gibbs random fields is thus
that their specific structure allows for a sufficient statistic vector that runs across models and
therefore leads to an exact (when $\epsilon=0$) simulation from the posterior probabilities of the models.
Each Gibbs random field model has its own sufficient statistic $\eta_m(\cdot)$ and
\cite{grelaud:marin:robert:rodolphe:tally:2009} exposed the fact that the vector of statistics
$\feta(\cdot)=(\eta_1(\cdot),\ldots,\eta_M(\cdot))$ is also sufficient for the joint parameter
$(\mathcal{M},\btheta_1,\ldots,\btheta_M)$. 

\cite{didelot:everitt:johansen:lawson:2011} point out that this specific property of Gibbs random fields can be
extended to any exponential family (hence to any setting with fixed-dimension sufficient statistics, see
\citealp{lehmann:casella:1998}). Their argument is that, by including all sufficient statistics and all
dominating measure statistics in an encompassing model, models under comparison are submodels of the
encompassing model.  The concatenation of those statistics is then jointly sufficient across models. While this
encompassing principle holds in full generality, in particular when comparing models that are already embedded,
we think it leads to an overly optimistic perspective about the merits of ABC for model choice: in practice,
most complex models do not enjoy sufficient statistics (if only because they are beyond exponential families). The
Gibbs case processed by \cite{grelaud:marin:robert:rodolphe:tally:2009} therefore happens to be one of the very
few realistic counterexamples. As demonstrated in the next section and in the normal example in SI, using
insufficient statistics is more than a mere loss of information. Looking at what happens in the limiting case
when one relies on a common model-wise sufficient statistic is a formal but useful study since it brings light
on the potentially huge discrepancy between the ABC-based and the true Bayes factors.  To develop a solution to
the problem in the formal case of the exponential families does not help in understanding the discrepancy for
non-exponential models.

\subsection{Arbitrary ratios}\label{Siouxie}

The difficulty with the discrepancy between $B_{12}(\by)$ and $B^{\feta}_{12}(\by)$ is that this discrepancy is
impossible to evaluate in a general setting, while there is no reason to expect a reasonable agreement between
both quantities. A first illustration was produced by \cite{marin:pudlo:robert:ryder:2011} in the case of
MA$(q)$ models.

\begin{figure}
\includegraphics[width=.4\textwidth]{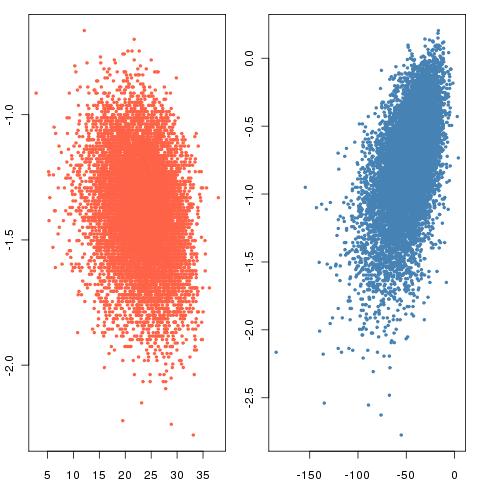}
\caption{\label{fig:poisneg}Comparison between the true log-Bayes factor {\em (first axis)} for the comparison of a Poisson
model versus a negative binomial model and of the log-Bayes factor based on the sufficient
statistic $\sum_i y_i$ {\em (second axis)}, for Poisson {\em (left)} and negative binomial {\em (left)}
samples of size $n=50$, based on $T=10^4$ replications}
\end{figure}

A simple illustration of the discrepancy due to the use of a model-wise sufficient statistic is a
a sample $\by=(y_1,\ldots,y_n)$ that could come either from a Poisson $\mathcal{P}(\lambda)$ distribution or
from a geometric $\mathcal{G}(p)$ distribution, already introduced in
\cite{grelaud:marin:robert:rodolphe:tally:2009} as a counter-example to Gibbs random fields and later
reprocessed in \cite{didelot:everitt:johansen:lawson:2011} to support their sufficiency argument.  In this
case, the sum $S=\sum_{i=1}^n y_i=\feta(\by)$ is a sufficient statistic for both models but not across models. The
distribution of the sample given $S$ is a multinomial $\mathcal{M}(S,1/n,\ldots,1/n)$ distribution when the
data is Poisson, 
while it is the uniform distribution over the $\by$'s such that $\sum_i y_i=S$
in the geometric case, since $S$ is then a negative binomial $\mathcal{N}eg(n,p)$ variable. The discrepancy
ratio is therefore
$$
{g_1(\by)}/{g_2(\by)} = {n+S-1\choose S}\,{S!\,n^{-S}\big/\prod_i y_i!} \,.
$$
When simulating $n$ Poisson or geometric variables and using prior distributions as exponential 
$\lambda \sim \mathcal{E}(1)$ and uniform $p\sim\mathcal{U}(0,1)$ on the parameters of
the respective models, the exact Bayes factor is available and the distribution of the
discrepancy is therefore available. Fig \ref{fig:poisneg} gives the range of $B_{12}(\by)$ versus
$B^{\feta}_{12}(\by)$, showing that $B^{\feta}_{12}(\by)$ is then unrelated with
$B_{12}(\by)$: the values produced by both approaches have nothing in common. As noted above, the
approximation $B^{\feta}_{12}(\by)$ based on the sufficient statistic $S$ is producing figures of the magnitude
of a {\em single} observation, while the true Bayes factor is of the order of the sample size.

The discrepancy between both Bayes factors is in fact increasing with the sample size, as shown by the
following result:

\begin{lemma}
Consider model selection between model 1: $\mathcal{P}(\lambda)$ with prior
distribution $\pi_1(\lambda)$ equal to an exponential $\mathcal{E}(1)$ distribution and
model 2: $\mathcal{G}(p)$ with a uniform prior distribution $\pi_2$ when the observed data $\by$ consists of
iid observations with expectation $\mathbb{E}[y_i] = \theta_0 > 0$. Then $S(\by) = \sum_{i=1}^n y_i$ is the
minimal sufficient statistic for both models and the Bayes factor based on the sufficient statistic $S(\by)$,
$B^{\feta}_{12}(\by)$, satisfies
$$ 
\lim_{n \rightarrow \infty} B^{\feta}_{12}(\by)= \theta_0^{-1} (\theta_0+1)^2 e^{-\theta_0}\quad \mbox{a.s.}
$$
\end{lemma}

Therefore, the Bayes factor based on the statistic $S(\by)$ is \emph{not} consistent; it converges to a
non-zero, finite value.

In this specific setting, \cite{didelot:everitt:johansen:lawson:2011} show that adding $P=\prod_i y_i!$ to the
$S$ creates a statistic $(S,P)$ that is sufficient across both models. While this is mathematically correct,
it does not provide further understanding of the behaviour of ABC-model choice in realistic
settings: outside formal examples as the one above and well-structured if complex exponential families like
Gibbs random fields, it is not possible to devise completion mechanisms that ensure sufficiency across models,
or even select well-discriminating statistics. It is therefore more fruitful to study and detect the diverging
behaviour of the ABC approximation as given, rather than attempting at solving the problem in a specific and
formal case.

\subsection{Population genetics} 

We recall that ABC has first been introduced by population geneticists
\citep{beaumont:zhang:balding:2002,pritchard:seielstad:perez:feldman:1999} for statistical inference about the
evolutionary history of species, because no likelihood-based approach existed apart from very simple and hence
unrealistic situations. This approach has since been used in an increasing number of biological studies
\citep{estoup:etal:2004,estoup:clegg:2003,fagundes:etal:2007}, most of them including model choice. It is
therefore crucial to get insights in the validity of such studies, particularly when they deal with species of
economical or ecological importance (see, e.g., \citealp{lombaert:etal:2010}). To this end, we need to compare
ABC estimates of posterior probabilities to reliable likelihood-based estimates.  Combining different modules
based on \cite{stephens:donnelly:2000}, it is possible to approximate the likelihood of population genetic data
through importance sampling (IS) even in complex scenarios.  In order to evaluate the potential discrepancy
between ABC-based and likelihood-based posterior probabilities of evolutionary scenarios, we designed two
experiments based on simulated data with limited information content, so that the choice between scenarios is
problematic. This setting thus provides a wide enough set of intermediate values of model posterior
probabilities, in order to better evaluate the divergence between ABC and likelihood estimates.  

In the first experiment, we consider two populations (1 and 2) having diverged at a fixed time in the past and
a third population (3) having diverged from one of those two populations (scenarios 1 and 2 respectively).
Times are set to 60 generations for the first divergence and to 30 generations for the second divergence. One
hundred pseudo observed datasets have been simulated, represented by 15 diploid individuals per population
genotyped at five independent microsatellite loci.  These loci are assumed to evolve according to the strict
Stepwise Mutation model (SMM), i.e.~when a mutation occurs, the number of repeats of the mutated gene increases
or decreases by one unit with equal probability. The mutation rate, common to all five loci, has been set to
$0.005$ and effective population sizes to 30. In this experiment, both scenarios have a single parameter: the
effective population size, assumed to be identical for all three populations. We chose a uniform prior
$U[2,150]$ for this parameter (the true value being 30). The IS algorithm was performed using 100 coalescent
trees per particle.  The marginal likelihood of both scenarios has been computed for the same set of 1000
particles and they provide the posterior probability of each scenario. The ABC computations have been performed
with DIYABC \citep{cornuet:santos:beaumont:etal:2008}. A reference table of 2 million datasets has been
simulated using 24 usual summary statistics (provided in Table S1) and the posterior probability of each
scenario has been estimated 
as their proportion in the 500 simulated datasets closest to the pseudo observed one. This population genetic
setting does not allow for a choice of sufficient statistics, even at the model level.

\begin{table*}
\noindent{\normalsize {\bf Tab.~S1}}
{\small Summary statistics used in the population genetic experiments, the Subset column corresponding to the
ABC
operated with 15 summary statistics and the last three statistics being only used in this reduced collection}
{\footnotesize
\begin{tabular*}{\hsize}{@{\extracolsep{\fill}}lll}
Name &Subset &Definition\\[3pt]
\hline\\
NAL1 & yes & average number of alleles in population 1\cr
NAL2 & yes & average number of alleles in population 2\cr
NAL3 & yes & average number of alleles in population 3\cr
HET1 & yes & average heterozygothy n population 1\cr
HET2 & yes & average heterozygothy n population 2\cr
HET3 & yes & average heterozygothy n population 3\cr
VAR1 & yes & average variance of the allele size in population 1\cr
VAR2 & yes & average variance of the allele size in population 2\cr
VAR3 & yes & average variance of the allele size in population 3\cr
MGW1 & no  & Garza-Williamson M in population 1\cr
MGW2 & no  & Garza-Williamson M in population 2\cr
MGW3 & no  & Garza-Williamson M in population 3\cr
FST1 & no  & average FST in population 1\cr
FST2 & no  & average FST in population 2\cr
FST3 & no  & average FST in population 3\cr
LIK12& no  & probability that sample 1 is from population 1\cr
LIK13& no  & probability that sample 1 is from population 3\cr
LIK21& no  & probability that sample 2 is from population 1\cr
LIK23& no  & probability that sample 2 is from population 3\cr
LIK31& no  & probability that sample 3 is from population 1\cr
LIK32& no  & probability that sample 3 is from population 2\cr
DAS12& yes & shared allele distance between populations 1 and 2\cr
DAS13& yes & shared allele distance between populations 1 and 3\cr
DAS23& yes & shared allele distance between populations 2 and 3\cr
DM212& yes & distance $(\delta\mu)^2$ between populations 1 and 2\cr
DM213& yes & distance $(\delta\mu)^2$ between populations 1 and 3\cr
DM223& yes & distance $(\delta\mu)^2$ between populations 2 and 2\\[3pt]
\hline
\end{tabular*}
}
\end{table*}

The second experiment also opposes two scenarios including three populations, two of them having diverged 100
generations ago and the third one resulting of a recent admixture between the first two populations (scenario
1) or simply diverging from population 1 (scenario 2) at the same time of 5 generations in the past. In
scenario 1, the admixture rate is $0.7$ from population 1. Pseudo observed datasets (100) of the same size as
in experiment 1 (15 diploid individuals per population, 5 independent microsatellite loci) have been generated
for an effective population size of 1000 and mutation rates of $0.0005$. In contrast with experiment 1,
analyses included the following 6 parameters (provided with corresponding priors): admixture rate
($U[0.1,0.9]$), three effective population sizes ($U[200,2000]$), the time of admixture/second divergence
($U[1,10]$) and the time of the first divergence ($U[50,500]$). To account for an higher complexity in the
scenarios, the IS algorithm was performed with 10,000 coalescent trees per particle. Apart from this change,
both ABC and likelihood analyses have been performed in the same way as experiment 1.  

\begin{figure}
\centering
\includegraphics[width=.29\textwidth]{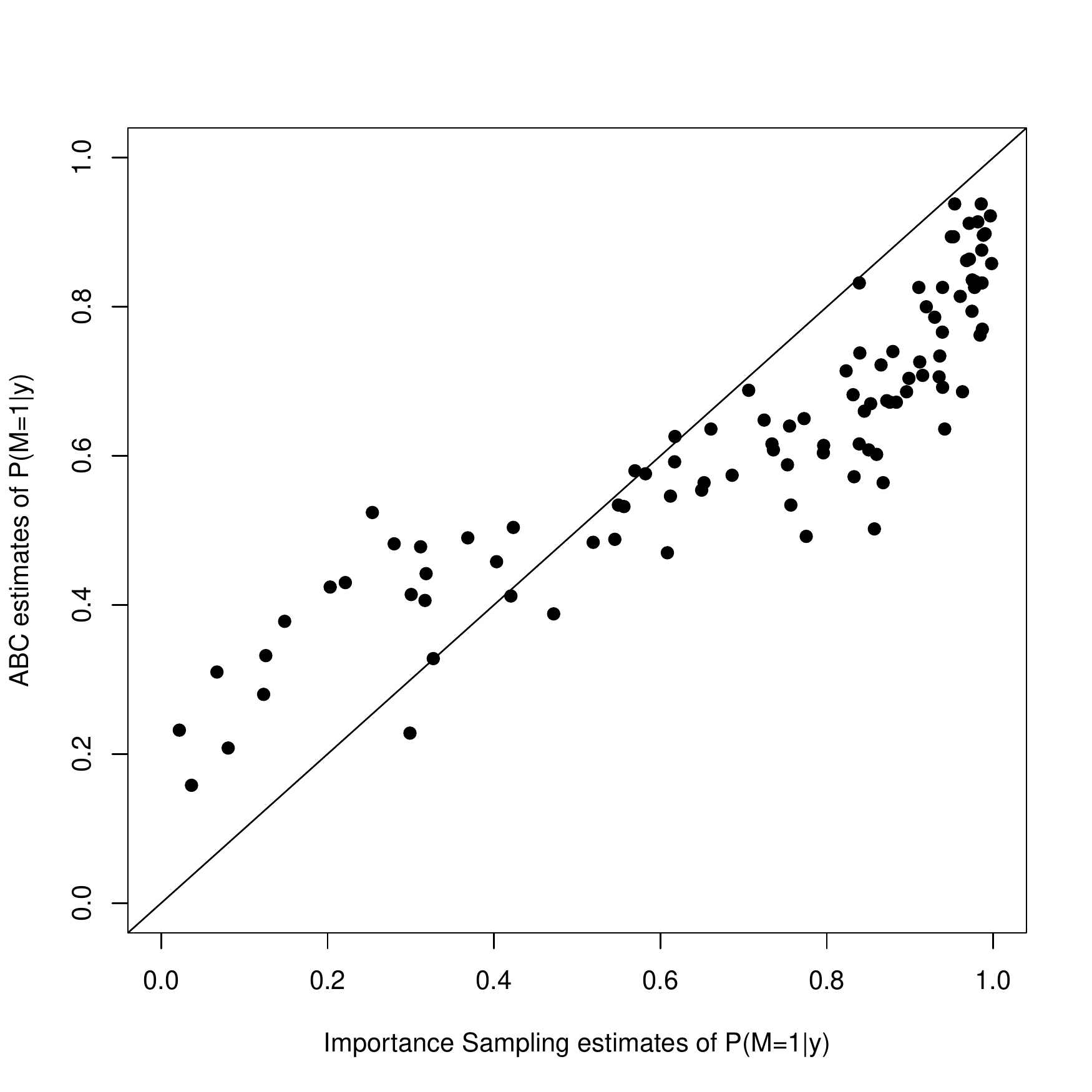}
\caption{\label{fig:res11}Comparison of IS and ABC estimates of the posterior probability of scenario
1 in the first population genetic experiment, using 24 summary statistics}
\end{figure}

\begin{figure}
\centering
\includegraphics[width=.29\textwidth]{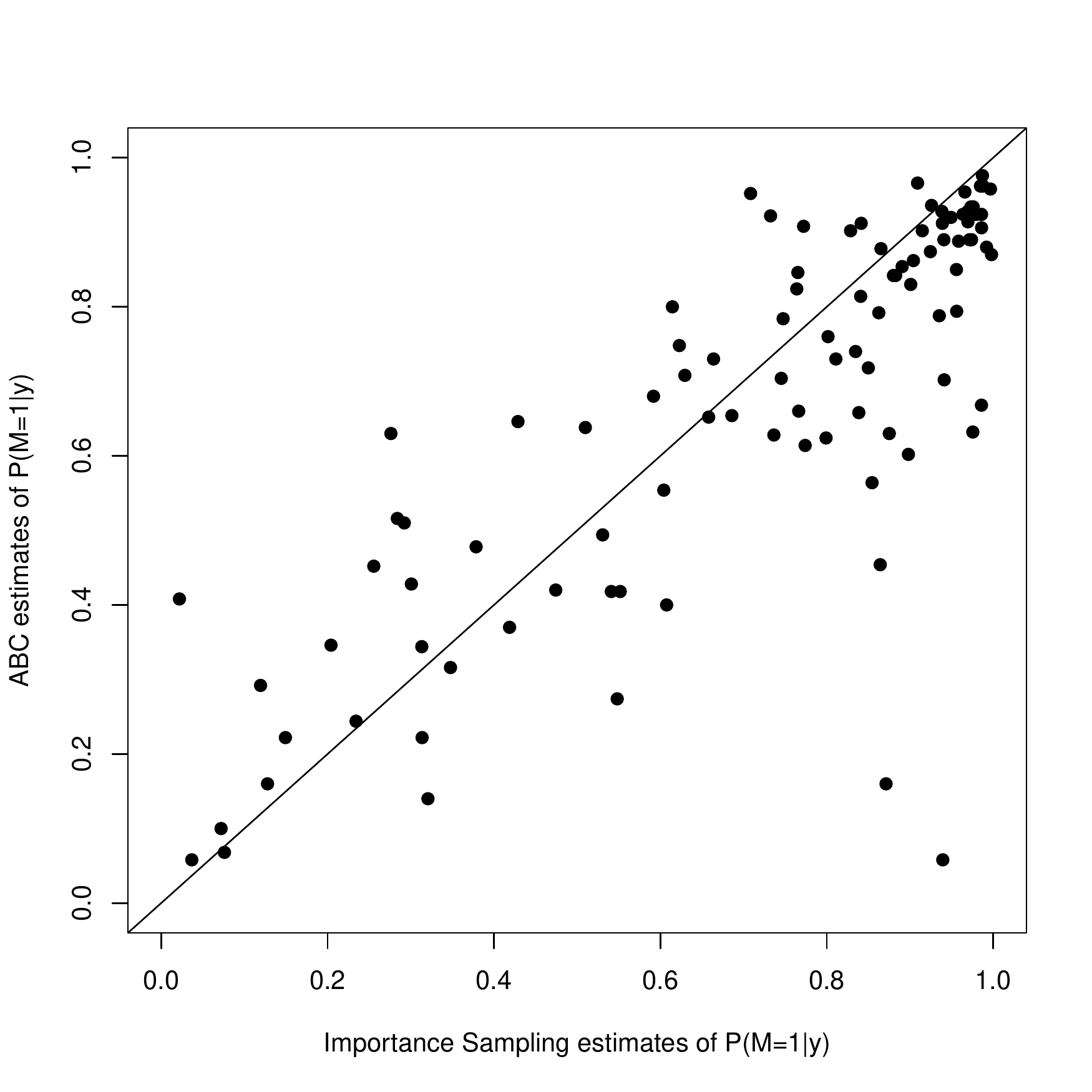}
\caption{\label{fig:res12}Same caption as Fig \ref{fig:res11} when using 15 summary statistics}
\end{figure}

\begin{figure}
\centering
\includegraphics[width=.29\textwidth]{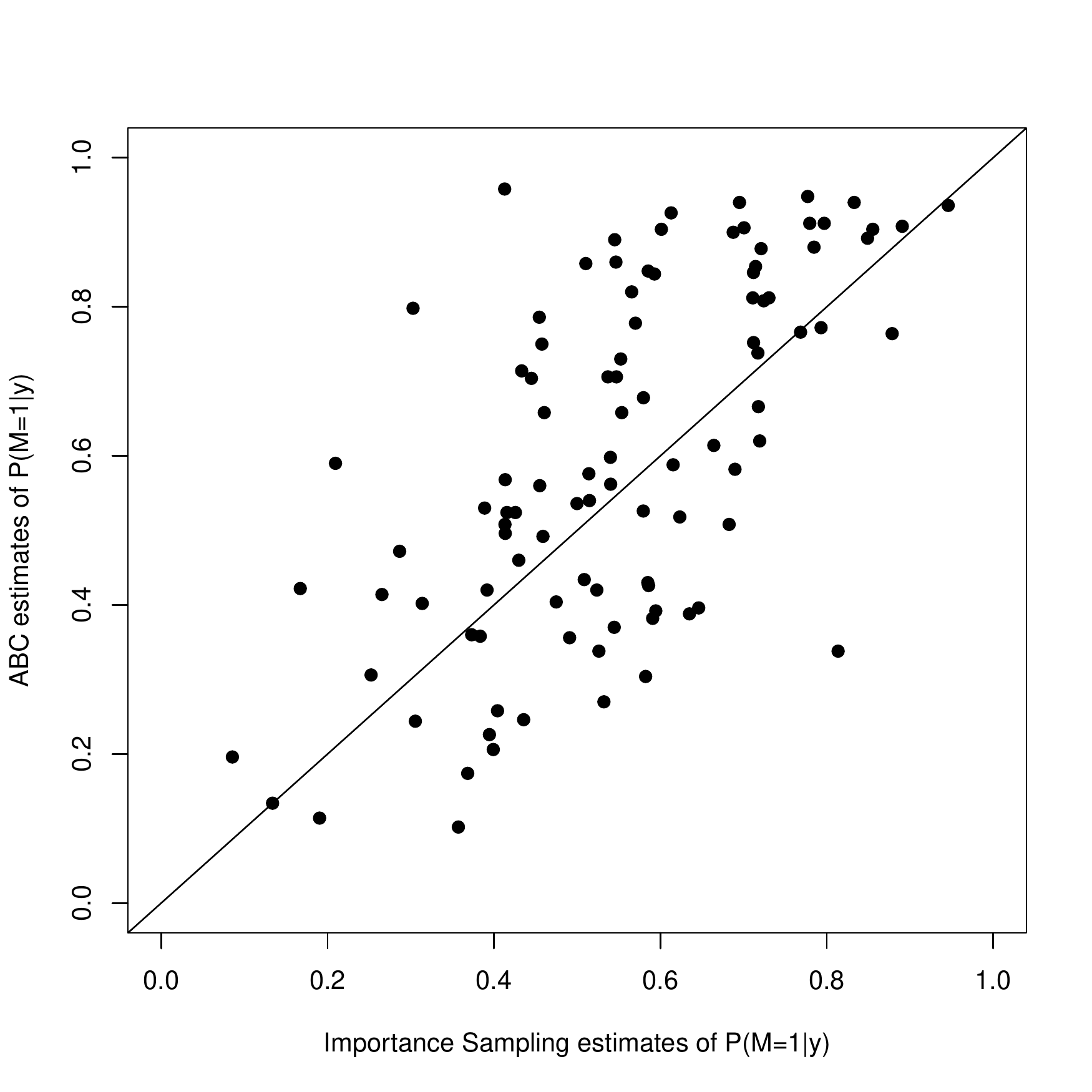}
\caption{\label{fig:res2}Comparison of IS and ABC estimates of the posterior probability of scenario
1 in the second population genetic experiment}
\end{figure}

Fig \ref{fig:res11} shows a reasonable fit between the exact posterior probability of model 1 (evaluated by
IS) and the ABC approximation in the first experiment on most of the 100 simulated datasets, even though the
ABC approximation is biased towards $0.5$. When using $0.5$ as the decision boundary between
model 1 and model 2, there is hardly any discrepancy between both approaches, demonstrating that model choice
based on ABC can be trusted in this case. Fig \ref{fig:res12} considers the same setting when moving from
24 to 15 summary statistics (given in Table S1): the fit somehow degrades. In particular, the
number of opposite conclusions in the model choice moves to $12\%$.  In the more complex setting of the second
experiment, the discrepancy worsens, as shown on Fig \ref{fig:res2}. The number of opposite conclusions
reaches $26\%$ and the fit between both versions of the posterior probabilities is considerably degraded,
with a correlation coefficient of $0.643$.

\begin{figure}
\includegraphics[width=.4\textwidth]{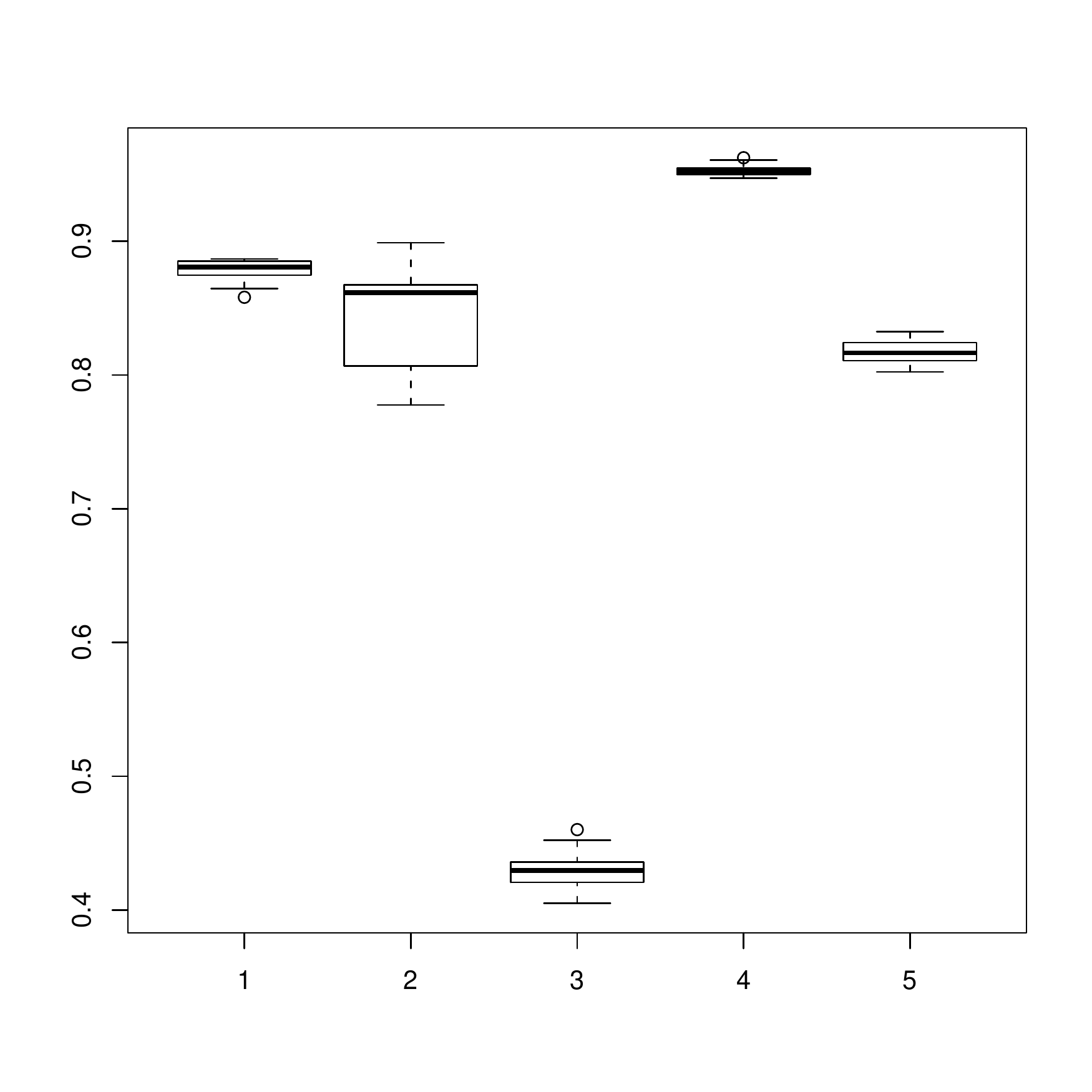}
\caption{\label{fig:repeat1} Boxplots of the posterior probabilities evaluated over 10 independent Monte Carlo
evaluations, for five independent simulated datasets in the first population genetic experiment}
\end{figure}

\begin{figure}
\includegraphics[width=.4\textwidth]{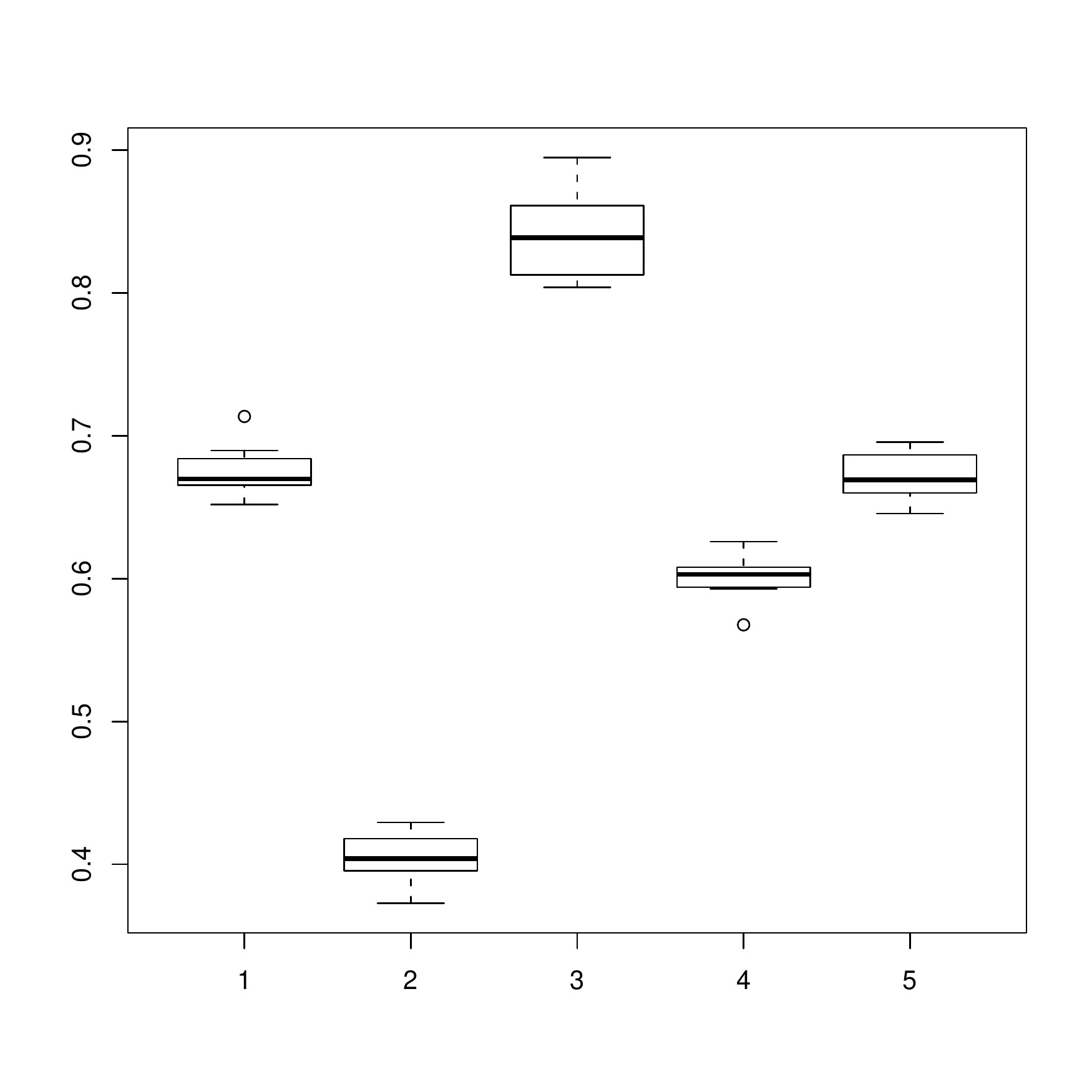}
\caption{\label{fig:repeat2} Boxplots of the posterior probabilities evaluated over 10 independent Monte Carlo
evaluations, for five independent simulated datasets in the second population genetic experiment}
\end{figure}

\begin{figure}
\includegraphics[width=.3\textwidth]{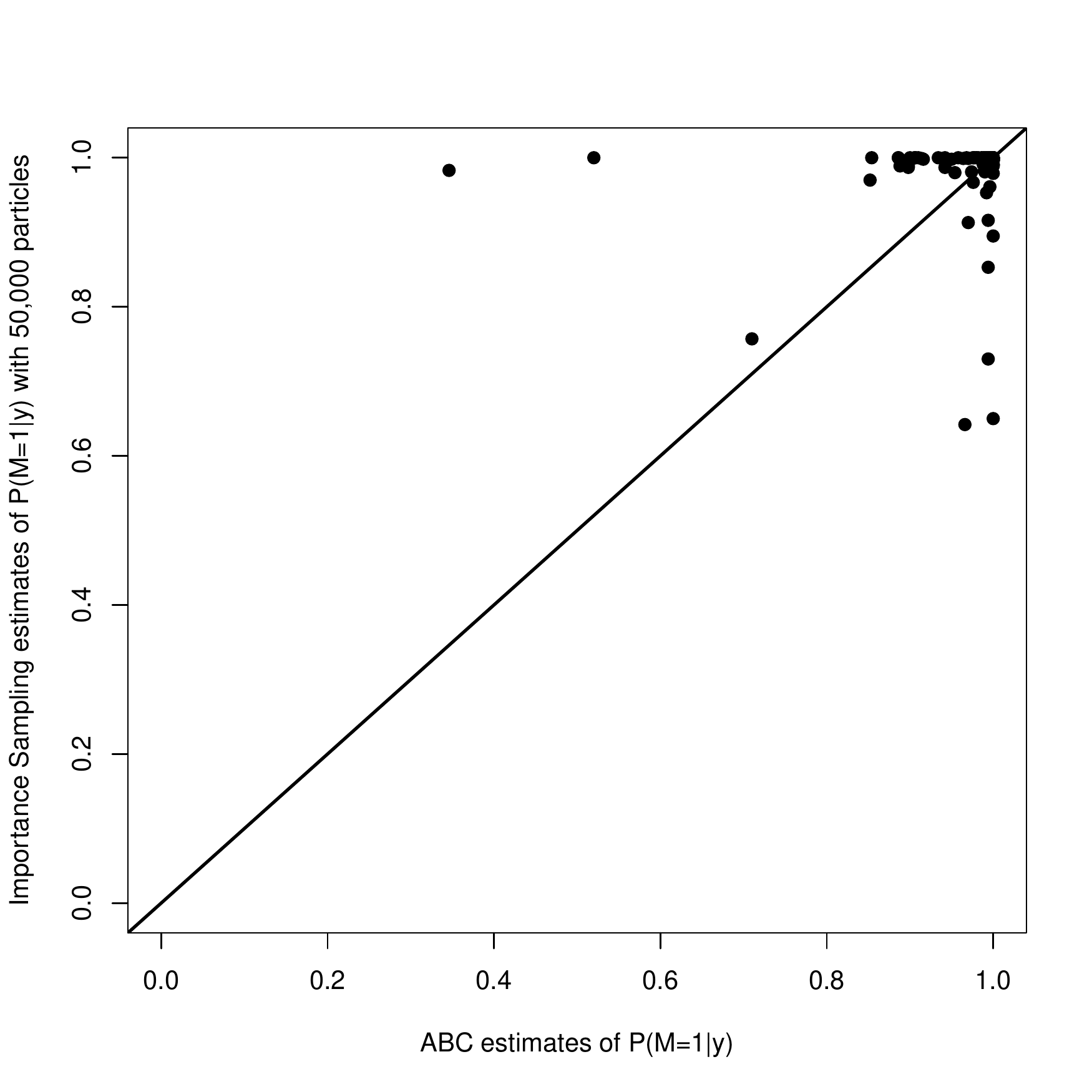}
\caption{\label{fig:resml2} Comparison between two approximations of the posterior probabilities of scenario 1
based on importance sampling with 50,000 particles {\em (first axis)} and ABC {\em (second axis)} for the
larger population genetic experiment}
\end{figure}

The validity of the importance sampling approximation can obviously be questioned in both experiments, however
Fig \ref{fig:repeat1} and Fig \ref{fig:repeat2} display a strong stability of the posterior probability IS approximation across 10
independent runs for 5 different datasets and gives proper confidence in this approach. 
Increasing the number of loci to 50 and the sample size to 100 individuals per population (see SI)
leads to posterior probabilities of the true scenario overwhelmingly close to one (Fig \ref{fig:resml2}),
thus bluring the distinction between ABC and likelihood based estimates but also reassuring on the ability of
ABC to provide the right choice of model with a higher information content of the data. Actually, we note that,
for this experiment, all ABC-based decisions conclude in favour of the correct model.

\section{Discussion}\label{RollingStones}

Since its introduction by \cite{tavare:balding:griffith:donnelly:1997} and
\cite{pritchard:seielstad:perez:feldman:1999}, ABC has been extensively used in several areas involving complex
likelihoods, primarily in population genetics, both for point estimation and testing of hypotheses. In
realistic settings, with the exception of Gibbs random fields, which satisfy a resilience property with respect
to their sufficient statistics, the conclusions drawn on model comparison cannot be trusted {\em per se} but
require further simulations analyses as to the pertinence of the (ABC) Bayes factor based on the summary
statistics.  This paper has examined in details only the case when the summary statistics are sufficient for
both models, while practical situations imply the use of insufficient statistics.  The rapidly increasing
number of applications estimating posterior probabilities by ABC indicates a clear need for further
evaluations of the worth of those estimations. 

Further research is needed for producing trustworthy approximations to the posterior probabilities of models.
At this stage, unless the whole data is involved in the ABC approximation as in \cite{sousa:etal:2009}, our
conclusion on ABC-based model choice is to exploit the approximations in an exploratory manner as measures of
discrepancies rather than genuine posterior probabilities. This direction relates with the analyses found in
\cite{ratmann:andrieu:wiujf:richardson:2009}.  Furthermore, a version
of this exploratory analysis is already provided in the DIY-ABC software of
\citep{cornuet:santos:beaumont:etal:2008}.  An option in this software allows for the computation of a Monte
Carlo evaluation of false allocation rates resulting from using the ABC posterior probabilities in selecting a
model as the most likely.  For instance, in the setting of both our population genetic experiments, DIY-ABC
gives false allocation rates equal to $20\%$ (under scenarios 1 and 2) and $14.5\%$ and $12.5\%$ (under
scenarios 1 and 2), respectively. This evaluation obviously shifts away from the performances of ABC as an
approximation to the posterior probability towards the performances of the whole Bayesian apparatus for
selecting a model, but this nonetheless represents a useful and manageable quality assessment for
practitioners.


\begin{acknowledgments}
The first three authors' work has been partly supported by Agence Nationale de la Recherche via the
2009--2013 project EMILE. They are grateful to the reviewers and to Michael Stumpf for their comments. 
Computations were performed on the INRA CBGP and MIGALE clusters.
\end{acknowledgments}

\small

\appendix[SI Results]

\subsection{A normal illustration} The following reproduces the Poisson geometric illustration in a normal model.
If we look at a fully normal $\mathcal{N}(\mu,\sigma^2)$ setting, we have
$$
f(\by|\mu) \propto \exp\left\{-n\sigma^{-2}(\bar y - \mu)^2/2 -\sigma^{-2} \sum_{i=1}^n (y_i-\bar y)^2 /2\right\} \sigma^{-n}
$$
hence
$$
f(\by | \bar y) \propto \exp\left\{-\sigma^{-2} \sum_{i=1}^n (y_i-\bar y)^2 /2\right\} \sigma^{-n} \mathbb{I}_{\sum y_i = n\bar y}\,.
$$
If we reparameterise the observations into $\bu = (y_1-\bar y,\ldots,y_{n-1}-\bar y,\bar y)$, we do get
\begin{eqnarray*}
f(\bu|\mu) &\propto&  \sigma^{-n}\, \exp\left\{-n\sigma^{-2}(\bar y - \mu)^2/2 \right\}\\
    &    & \times\exp\left\{-\sigma^{-2} \sum_{i=1}^{n-1} u_i^2/2 - 
             \sigma^{-2} \left[ \sum_{i=1}^{n-1} u_i \right]^2  \big/2 \right\}
\end{eqnarray*}
since the Jacobian is $1$. Hence
$$
f(\bu|\bar y) \propto \exp\left\{-\sigma^{-2} \sum_{i=1}^{n-1} u_i^2/2 - 
    \sigma^{-2} \left[ \sum_{i=1}^{n-1} u_i \right]^2  /2 \right\} \sigma^{-n} 
$$
Considering both models
$$
y_1,\ldots,y_n \stackrel{\text{iid}}{\sim} \mathcal{N}(\mu,\sigma_1^2)\quad\text{ and }\quad
y_1,\ldots,y_n \stackrel{\text{iid}}{\sim} \mathcal{N}(\mu,\sigma_2^2)\,,
$$
the discrepancy ratio is then given by
$$
\dfrac{\sigma_2^{n-1}}{\sigma_1^{n-1}}\,
\exp\left\{\dfrac{\sigma_2^{-2}-\sigma_1^{-2}}{2}\left(\sum_{i=1}^{n-1} (y_i-\bar y)^2+ \left[ \sum_{i=1}^{n-1}
(y_i-\bar y) \right]^2  \right)\right\} 
$$
and is connected with the lack of consistency of the Bayes factor:

\begin{lemma}
Consider model selection between model 1: $\mathcal{N}(\mu,\sigma_1^2)$ and model 2:
$\mathcal{N}(\mu,\sigma_2^2)$, $\sigma_1$ and $\sigma_2$ being given, with prior distributions
$\pi_1(\mu)=\pi_2(\mu)$ equal to a $\mathcal{N}(0,a^2)$ distribution and when the observed data $\by$ consists
of iid observations with finite mean and variance.  Then $S(\by) = \sum_{i=1}^n y_i$ is the minimal sufficient
statistic for both models and the Bayes factor based on the sufficient statistic $S(\by)$,
$B^{\feta}_{12}(\by)$, satisfies
$$
\lim_{n \rightarrow \infty} B^{\feta}_{12}(\by)= 1
\quad\mbox{a.s.}
$$
\end{lemma}

\begin{figure}
\includegraphics[width=.4\textwidth]{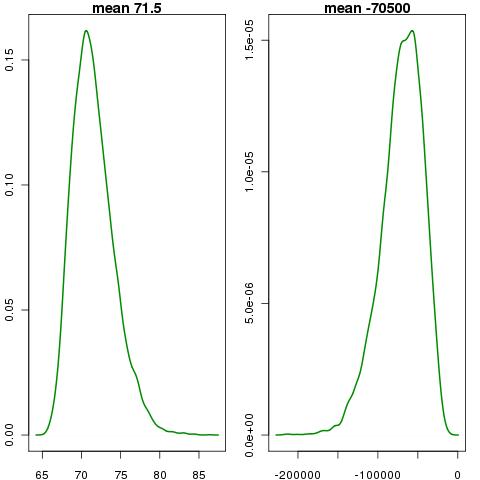}
\caption{\label{fig:twonormal}Empirical distributions of the log discrepancy $\log g_1(\by)/ g_2(\by)$ for 
datasets of size $n=15$ simulated from $\mathcal{N}(\mu,\sigma_1^2)$ {\em (left)} and $\mathcal{N}(\mu,\sigma_2^2)$ {\em (right)}
distributions when $\sigma_1=0.1$ and $\sigma_2=10$, based on $10^4$ replications and a flat prior}
\end{figure}

Fig \ref{fig:twonormal} illustrates the behaviour of the discrepancy ratio when $\sigma_1=0.1$ and
$\sigma_2=10$, for datasets of size $n=15$ simulated according to both models. The discrepancy (expressed on a
log scale) is once again dramatic, in concordance with the above lemma.

If we now turn to an alternative choice of sufficient statistic, using the pair $(\bar y,S^2)$ with
$$
S^2=\sum_{i=1}^n (y_i-\bar y)^2\,,
$$
we follow the solution of \cite{didelot:everitt:johansen:lawson:2011}. Using a conjugate prior
$\mu\sim\mathcal{N}(0,a^2)$, the true Bayes factor is equal to the Bayes factor based on the corresponding
distributions of the pair $(\bar y,S^2)$ in the respective models. Therefore, with sufficient computing power,
the ABC approximation to the Bayes factor can be brought arbitrarily close to the true Bayes factor.  However,
this coincidence does not bring any intuition on the behaviour of the ABC approximations in realistic settings.

\subsection{Larger experiment}

We also considered a more informative population genetic experiment with the same scenarios (1 and 2) as in the
second experiment. One hundred datasets were simulated under scenario 1 with 3 populations, i.e.~6 parameters.
We take 100 diploid individuals per population, 50 loci per individual. This thus corresponds to 300 genotypes
per dataset. The IS algorithm was performed using 100 coalescent trees per particle. The marginal likelihood of
both scenarios has been computed for the same set for both 1000 particles (IS1) and 50,000 particles (IS2). A
national cluster of 376 processors (including 336 Quad Core processors) was used for this massive experiment
(which required more than 12 calendar days for the importance sampling part).

The confidence about the IS approximation can be assessed on Fig \ref{fig:resml2}, which shows that both runs
most always provide the same numerical value, which almost uniformly is very close to one.  This makes the fit
of the ABC approximation to the true value harder to assess, even though we can spot a trend towards
under-estimation. Furthermore, they almost all lead to correctly select model 1.

\end{article}
\end{document}